\documentclass{Interspeech2024}

\interspeechcameraready

\title{BiVocoder: A Bidirectional Neural Vocoder Integrating Feature Extraction and Waveform Generation}
\usepackage{graphicx}
\usepackage{subcaption} %
\usepackage{authblk}
\usepackage{multirow}
\usepackage{adjustbox}
\name{Hui-Peng}{Du}
\name{Ye-Xin}{Lu}
\name{Yang}{Ai}
\name{Zhen-Hua}{Ling}

\address{
  National Engineering Research Center of Speech and Language Information
Processing,\\
University of Science and Technology of China, Hefei, P. R. China}
\email{\{redmist, yxlu0102\}@mail.ustc.edu.cn, \{yangai, zhling\}@ustc.edu.cn}


\begin{document}

\maketitle

\begin{abstract}
This paper proposes a novel bidirectional neural vocoder, named BiVocoder, capable both of feature extraction and reverse waveform generation within the short-time Fourier transform (STFT) domain. 
For feature extraction, the BiVocoder takes amplitude and phase spectra derived from STFT as inputs, transforms them into long-frame-shift and low-dimensional features through convolutional neural networks. 
The extracted features are demonstrated suitable for direct prediction by acoustic models, supporting its application in text-to-speech (TTS) task.  
For waveform generation, the BiVocoder restores amplitude and phase spectra from the features by a symmetric network, followed by inverse STFT to reconstruct the speech waveform. 
Experimental results show that our proposed BiVocoder achieves better performance compared to some baseline vocoders, by comprehensively considering both synthesized speech quality and inference speed for both analysis-synthesis and TTS tasks. 
\end{abstract}

\section{Introduction}

Neural vocoders have made tremendous advancements in recent years, significantly impacting the quality of synthesized speech in various tasks such as text-to-speech (TTS), singing voice synthesis (SVS), voice conversion (VC), and speech bandwidth expansion (BWE). 
Reviewing the development of vocoders, earlier signal-processing-based bidirectional conventional vocoders like WORLD \cite{morise2016world} and STRAIGHT \cite{kawahara1999restructuring} simultaneously possess the functions of feature extraction and waveform generation. 
For example, the STRAIGHT can extract fundamental frequency (F0) and mel-cepstral coefficients from speech waveforms and resynthesize speech waveforms based on these features. 
However, the synthesized speech quality of these conventional vocoders is always unsatisfactory. 

With the advancement of deep learning, unidirectional neural vocoders have been proposed for waveform generation tasks (i.e., missing feature extraction function). 
Mainstream neural vocoders \cite{kong2020hifi, ai2023apnet, kaneko2022istftnet, siuzdak2023vocos} use mel spectrogram as input features, while there are also others such as neural source-filter vocoders \cite{yang2023fast, yoneyama2023source} that use F0 and other acoustic features. 
These features are directly extracted from the raw waveforms using digital signal processing (DSP) methods, but they discard the crucial phase information, limiting the precise phase prediction and higher-quality speech generation. 
Recent works, such as Autovocoder \cite{webber2023autovocoder}, propose to use neural networks to learn an acoustic feature without discarding phase, and further resynthesize waveforms from the learned features by differentiable DSP (DDSP) \cite{engel2019ddsp}. 
Therefore, Autovocoder is a bidirectional neural vocoder. 
However, the features extracted by Autovocoder has even higher dimensionality than traditional mel spectrogram, which didn't show significant advantages in terms of computational complexity. 
Moreover, Autovocoder still optimizes through mel spectrogram loss and generative adversarial network (GAN) \cite{goodfellow2014generative} loss defined on waveforms, neglecting explicitly phase optimization, which limits the quality of the synthesized speech.

 In this paper, we propose BiVocoder, a bidirectional neural vocoder that also utilizes DDSP to perform feature extraction and waveform generation. 
At the feature extraction stage, the feature extraction module employs ConvNeXt V2 \cite{woo2023ConvNeXt} as backbone to perform deep processing on both amplitude and phase spectra extracted from speech waveform via short-time Fourier transform (STFT). 
Subsequently, downsampling and dimension reduction are employed to encode a long-frame-shift and low-dimensional feature.
At the waveform generation stage, these extracted features are then restored to amplitude and phase spectra by a symmetrical waveform generation module, and high-quality speech waveforms are reconstructed through inverse STFT (iSTFT). 
The feature extraction and waveform generation modules are bridged by these extracted features. 
Inspired by \cite{ai2023neural}, adversarial training strategy incorporating multi-spectral-level losses are adopted to train the feature extraction and waveform generation modules jointly. 
This approach enables precise amplitude and phase prediction, as well as high-quality waveform reconstruction. 
The experiments demonstrate that the proposed BiVocoder is capable of achieving high synthesized speech quality in analysis-synthesis tasks. 
Furthermore, the features extracted by the feature extraction module of the BiVocoder are conducive to being acquired by acoustic models. 
Consequently, in TTS tasks, BiVocoder attains performance on par with existing TTS models that utilize mel spectrograms as bridged features. 


\begin{figure*}[t]
	\centering
	\includegraphics[width=\linewidth]{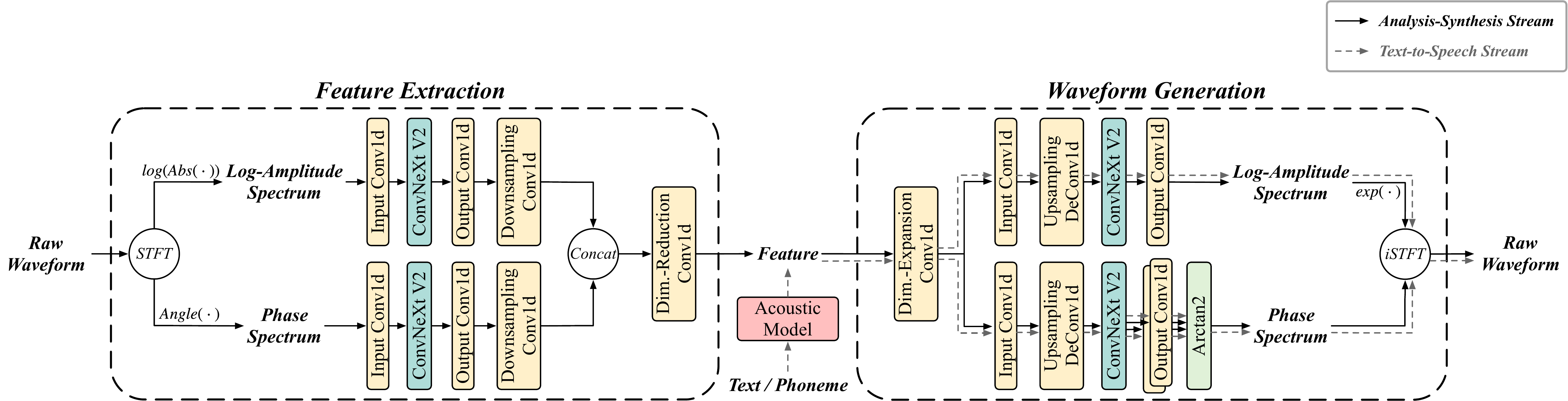}
	\caption{The architecture of BiVocoder and discriminators are omitted in the diagram. $ABS(\cdot)$ and $Angle(\cdot)$ denote amplitude and phase spectrum calculations. $Arctan2$ stands for two-arguement arc-tan function. Conv1d and DeConv1d represents 1D convolutional layer and 1D deconvolutional layer, respectively.}
	\label{fig:model}
\end{figure*}
\begin{figure}[t]
	\centering
	\includegraphics[width=\linewidth]{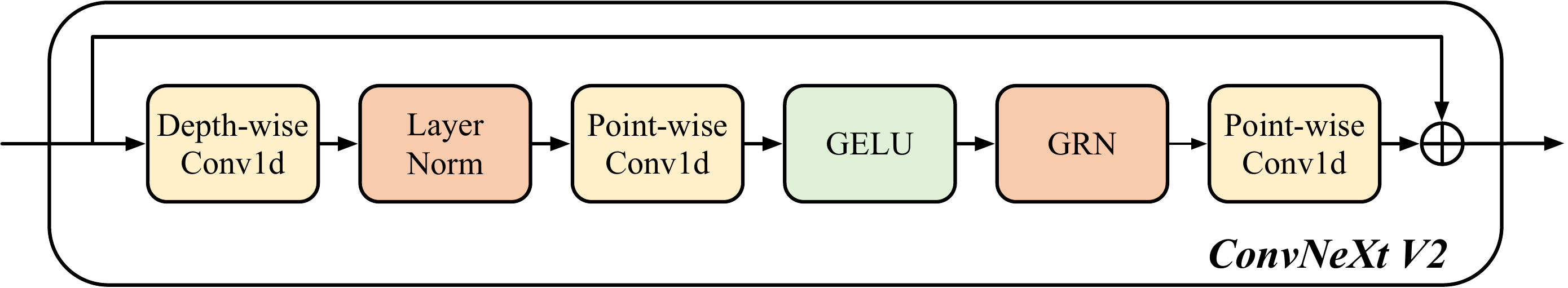}
	\caption{The architecture of the ConvNeXt V2 block, where GELU, and
		GRN represent Gaussian error linear unit, and global response
		normalization, respectively.}
	\label{fig:con}
\end{figure}

\section{Related Work}
\label{sec:realted}
 Based on the methods for feature extraction and waveform generation, we classify current vocoders into three categories. 
 \begin{itemize}
 	\item\textbf{Bidirectional conventional vocoder}. The bidirectional conventional vocoders, e.g., WORLD \cite{morise2016world} and STRAIGHT \cite{kawahara1999restructuring}, which uses traditional DSP methods to extract features (e.g., F0 and mel-cepstral coefficients) from input waveform and reconstruct the original waveform from these features. 
This type of vocoders has good versatility and can be directly processed for any data without the need for additional adaptation. 
However, their synthesized speech quality is poor compared to neural methods.
 	\item\textbf{Unidirectional neural vocoder}. The unidirectional neural vocoders \cite{kong2020hifi, ai2023apnet, oord2016wavenet, oord2018parallel, valin2019lpcnet} does not possess feature extraction capabilities. 
It can only take acoustic features (i.e., mel spectrogram) extracted by DSP methods as input for waveform generation. 
For example, HiFi-GAN \cite{kong2020hifi} is a fully convolutional neural network that directly predicts time-domain waveform from mel spectrogram. 
It achieves this by employing multiple deconvolutional layers to progressively upsample the input, matching the waveform's sampling rate. Our privious work, APNet \cite{ai2023apnet}, is also a fully convolutional model. 
It differs in that it simultaneously predicts the amplitude and phase spectra from input mel spectrogram rather than directly predicting waveforms. 
This approach effectively improves the generation efficiency. 
However, while this type of vocoders achieves high-quality synthesized speech, their upper limit in quality is constrained by the input features. 
Features lacking sufficient information inevitably impact the further improvement of performance in such vocoders. 
Recently, some studies \cite{loweimi2021speech, espic2017direct} also demonstrated that phase information, which is lost in the mel spectrogram, proves beneficial for waveform generation. 
 	\item\textbf{Bidirectional neural vocoder}. The bidirectional neural vocoder simultaneously achieves feature extraction and waveform generation through neural networks, compensating for the shortcomings of the unidirectional neural vocoders. Autovocoder \cite{webber2023autovocoder} is representative of this type of vocoders. 
It adopts an encoder-decoder architecture combined with DDSP to extract features from input waveform by an encoder network and then reconstruct the original waveform by a symmetrical decoder network. 
According to \cite{webber2023autovocoder}, the dimension of the features extracted by Autovocoder is even higher than that of the mel spectrogram used in unidirectional neural vocoders.  
Additionally, there is also a lack of validation regarding the application of these features to TTS, i.e., whether they are easily predictable by existing acoustic models. 
\end{itemize}

\section{Proposed Method}
\label{sec:method}

As demonstrated in Figure \ref{fig:model}, the architecture of BiVocoder can be divided into two main parts, i.e., feature extraction module and waveform generation module. 
In the feature extraction module, the input speech waveform undergoes with STFT, and the resulting amplitude and phase spectra are parallel processed to obtain a long-frame-shift and low-dimensional features. 
In the waveform generation module, the extracted features are processed to reconstruct the amplitude spectrum and phase spectrum in parallel, and subsequently undergoes iSTFT to reconstruct the raw speech waveform. 
Further insights of the model structure, training criteria, and the application of our model in the field of TTS are described as follows. 

\vspace{-1mm}
\subsection{Model Structure}
\vspace{-1mm}

The feature extraction and waveform generation in BiVocoder are mirror processes, thus the feature extraction module and the waveform generation module possess symmetrical structures, as shown in Figure \ref{fig:model}. 
Both modules are designed with a dual-branch architecture, couple amplitude and phase information through parallel branches into acoustic features, from which amplitude and phase are then decoupled. 
For these amplitude and phase branches within both modules, ConvNeXt V2 is employed as the backbone network because its fewer parameters and better modeling capabilities. 
Each ConvNeXt V2 network comprises multiple ConvNeXt V2 blocks as shown in Figure \ref{fig:con}, equipped with a convolutional layer featuring a large kernel designed to capture information from an expansive receptive field. 
Following normalization of each block's output through layer normalization, a 1$\times$1 pointwise convolution is applied to extract features in a high-dimensional space. 
These features undergo additional normalization via Gaussian error linear unit (GELU) activation \cite{hendrycks2016gaussian} and global residual normalization (GRN) \cite{woo2023ConvNeXt}, before being dimensionally reduced back to the input level using a 1$\times$1 convolution. 
Ultimately, the output of the ConvNeXt V2 block is integrated with the input through residual connections and forwarded to the subsequent layer. 
The features processed by the ConvNeXt V2 network are then fed into an output convolutional layer and a large-stride convolutional layer for downsampling. 
Finally, after concatenating the outputs from the two branches, a dimension-reducing convolution is used to yield the long-frame-shift and low-dimensional features integrated both amplitude and phase information. 

In the waveform generation module, the low-dimensional feature space is first expanded using a dimension-expanding convolutional layer. 
Then, the amplitude and phase branches are separated. 
In each branch, the input undergoes an input convolutional layer, followed by a deconvolutional layer for upsampling. 
Similar to feature extraction module, we also use ConvNeXt V2 blocks as the backbone network for two branches. 
The amplitude spectrum is obtained through an output convolutional layer, while for the phase spectrum prediction, we adopt a parallel spectrum estimation architecture as suggested in \cite{ai2023neural}. 

\vspace{-1mm}
\subsection{Training Criteria}
\vspace{-1mm}

 We adopt the GAN training strategy and utilize the hinge GAN loss function as delineated in \cite{siuzdak2023vocos, zeghidour2021soundstream}.  
 To enhance the discriminative capacity of our model, we incorporate both multi-period discriminators \cite{kong2020hifi} and multi-resolution discriminators \cite{jang2021univnet} into our training regimen. 
 Besides, to achieve precise spectral modeling, the amplitude spectrum loss, phase anti-wrapping loss, short-time complex spectrum loss and mel spectrogram proposed in \cite{ai2023apnet} are also used in the adversarial training process.
 
\vspace{-1mm}
\subsection{TTS Application}
\label{ssec:TTS}
\vspace{-1mm}

As depicted by the gray dashed line in Figure 1, when applying BiVocoder to the TTS task, the text or phoneme sequence first goes through an acoustic model to predict the features extracted by BiVocoder. 
Finally, the waveform generation module of the BiVocoder synthesizes the speech waveform from input predicted features. 
During the training phase of the acoustic model, the feature extraction module of BiVocoder provides training targets for the acoustic model. 

\section{Experiments Setup}
\label{sec:exp}

\vspace{-1mm}
\subsection{Dataset} 
\vspace{-1mm}

For the main experiment in Section \ref{sec: Analysis-synthesis task} and \ref{sec: TTS task}, we utilized the VCTK-0.92 dataset \cite{yamagishi2019cstr}. 
The VCTK dataset consists of speech utterances from 108 native English speakers, with a total duration of about 44 hours. 
We selected 2,937 utterances from 8 speakers as the test set. 
From 40,936 utterances from the remaining 100 speakers, we randomly selected 90\% as the training set and rest 10\% as the validation set. 
To assess vocoders' generalizability, we also conducted cross-dataset experiments in Section \ref{sec: Cross-dataests task} on the LJSpeech dataset \cite{ito2017}, only using 1,310 randomly selected speech samples for testing. 
All utterances were downsampled to 16 kHz for experiments. 

\vspace{-1mm}
\subsection{Implementation}
\vspace{-1mm}

In the proposed BiVocoder\footnote{Examples
	of generated speech can be found at demo page: https://redmist328.github.io/BiVcoder\_demo.}, the amplitude and phase spectra were extracted by STFT with frame length, frame shift, and FFT size of 20 ms, 2.5 ms, and 1024 respectively.  
For each module, the number of ConvNeXt v2 blocks was both set to 8. 
Except for the 1×1 convolution, the kernel size for other convolutions was 7. 
The downsampling/upsampling rate of these two module was 8. 
The resulting features had a frame shift of 20 ms and a dimensionality of 32 (i.e., long-frame-shift and low-dimensional), facilitating storage and transmission. 
We trained the model using the AdamW optimizer \cite{kingma2014adam}  up to 2 million steps. 
During training, we randomly cropped the speech clips to 8000 samples and set the batch size to 16. 

\vspace{-1mm}
\subsection{Baselines}
\vspace{-1mm}
We compared our proposed BiVocoder with bidirectional conventional vocoder STRAIGHT \cite{kawahara1999restructuring}, unidirectional neural vocoder HiFi-GAN\footnote{https://github.com/jik876/hifi-gan.} \cite{kong2020hifi} and APNet\footnote{https://github.com/YangAi520/APNet.} \cite{ai2023apnet}, and bidirectional neural vocoder Autovocoder\footnote{https://github.com/hcy71o/autovocoder.} \cite{webber2023autovocoder}. 
Firstly, we adhered to the feature configurations as outlined in their original papers. 
For STRAIGHT, 41-dimensional mel-cepstral coefficients and F0 with frame shift of 5 ms were used. 
The 80-dimensional mel spectrogram was utilized by both HiFi-GAN and APNet, albeit with different frame shifts of 10 ms and 5 ms respectively. 
For Autovocoder, the frame shift and dimension of the features were 10 ms and 256, respectively. 
Compared to these baseline vocoders, the features extracted by BiVocoder had a longer frame shift and lower dimensionality (i.e., frame shift of 20 ms and dimensionality of 32). 
Then, for fair comparison, we also conducted experiments to reproduced the baseline vocoders (except STRAIGHT, using * for representation) using the feature configuration of BiVocoder. 

\begin{table*}[h!]\Huge
	\centering
	\caption{Objective evaluation results of STRAIGHT, HiFi-GAN, APNet, Autovocoder and proposed BiVocoder. The trainable vocoders are trained on the VCTK dataset. UTMOS (VCTK) and UTMOS (LJSpeech) demonstrate the UTMOS scores of compared vocoder testing on VCTK test set and LJSpeech test set, respectively. ``*'' indicates that the used features have the same frame shift and dimensionality as those extracted by BiVocoder. The \textbf{bold} and \underline{underline} numbers indicate the optimal and sub-optimal results, respectively.
}\label{tab1}
	\adjustbox{width=\textwidth}{
		\renewcommand{\arraystretch}{1.1}
		\begin{tabular}{l c c c c c c c}
			\hline
			\hline
			  & {SNR(dB)$\uparrow$} & {LAS-RMSE(dB)$\downarrow$}& {MCD(dB)$\downarrow$} &{F0-RMSE(cent)$\downarrow$}& {V/UV error(\%)$\downarrow$}& {UTMOS (VCTK)$\uparrow$}& {UTMOS (LJSpeech)$\uparrow$}\\
			\hline
			{Natural}& -&-&-&-&-&4.04&4.38\\
			\hline
			{STRAIGHT} & 0.11
			&10.08&2.16&35.19&\underline{3.66}&2.80 & 3.26\\
			\hline
			{HiFi-GAN} & 2.28&6.08&2.21&67.86&6.80&3.84&3.86\\
			{HiFi-GAN*} & 2.30&6.97&2.31&148.33&10.64&3.47&3.73\\
			\hline
			{APNet} & \underline{6.62}&\textbf{4.07}&\textbf{0.87}&\textbf{23.77}&\textbf{3.14}&\underline{3.85} &3.60\\
			
			{APNet*} & 1.29&10.17&2.70&131.90&15.48&2.14 &1.75\\
			\hline
			{Autovocoder} & -0.82&11.63&4.47&67.79&7.91&3.42&\underline{4.09}\\
			{Autovocoder*} & -10.95&13.12&8.28&555.97&21.61&1.39&1.31\\
			\hline
			{Bivovoder} & \textbf{6.98}&\underline{5.53}&\underline{1.54}&\underline{23.79}&4.64&\textbf{4.06}&\textbf{4.31}\\

			\hline
			\hline
	\end{tabular}}
\end{table*}

\vspace{-1mm}
\subsection{Evaluation metrics}
\vspace{-1mm}

 In this study, we employed five objective metrics to assess
the quality of synthesized speech, as utilized in our previous work \cite{ai2023apnet}. 
These metrics include signal-to-noise ratio (SNR), root mean square error (RMSE) of logarithmic amplitude spectra (LAS-RMSE), mel-cepstrum distortion (MCD), RMSE of F0 (F0-RMSE), and voiced/unvoiced (V/UV) error. 
For the analysis-synthesis task, we also utilized the highly effective UTMOS tool\footnote{https://github.com/sarulab-speech/UTMOS22.} \cite{saeki2022utmos} for objective mean opinion score (MOS) prediction.
Additionally, the real-time factor (RTF), which is defined as the seconds required to generate one second of speech using a single NVIDIA 2080Ti GPU or a single Intel Xeon E5-2620 CPU core, was used as an objective metric to evaluate the efficiency of the waveform generation process. 
The feature extraction process does not involve RTF calculation because the differences among these vocoders are too significant. 

To assess the subjective quality of different vocoders applied to TTS task, we conducted mean opinion score (MOS) tests. 
Each MOS test involved 30 test utterances synthesized by these vocoders, alongside natural utterances. 
We gathered feedback from at least 25 native English listeners on the Amazon Mechanical Turk (AMT) crowdsourcing platform. 
Listeners were asked to rate the naturalness on a scale of 1 to 5, with a score interval of 0.5.

\section{Results and Analysis}

\subsection{Evaluations on Analysis-Synthesis Task}
\label{sec: Analysis-synthesis task}

For the analysis-synthesis task, we first analyzed the comparative results of the proposed BiVocoder and other baseline vocoders under the original configurations. 
As shown in Table \ref{tab1}, the BiVocoder outperformed the bidirectional conventional vocoder (i.e., STRAIGHT) and bidirectional neural vocoder (i.e., Autovocoder) on all metrics. 
However, compared with unidirectional neural vocoders, the APNet demonstrated more prominent results, especially on amplitude-related metrics, e.g., LAS-RMSE and MCD. 
One possible reason is that APNet used mel spectrograms as input, and it explicitly modeled amplitudes during waveform generation. 
While BiVocoder's features encompassed both amplitude and phase information, resulting in higher overall synthesized speech quality according to UTMOS results. 
For more evidences, we conducted ABX preference tests on AMT to compare the subjective quality of synthesized speech of APNet and BiVocoder. 
The preference scores for APNet, BiVocoder and neutrality were 30.3\%, 44.6\% and 25.1\%, respectiverly ($p <$ 0.01 of a $t$ test). 
This indicates that in terms of perception, BiVocoder was significantly better than APNet. 
Therefore, despite the longer frame shift and lower dimensionality of the features extracted by BiVocoder compared to other features, the BiVocoder still achieved the highest synthesized speech quality, confirming the powerful modeling capability of the proposed model. 

When using the same feature configuration as the BiVocoder, both the DDSP-based APNet and Autovocoder experienced a severe decline in performance. 
This could be because the prediction of spectra (especially phase spectra) is sensitive to frame shifts \cite{ai2023long}. 
However, the issue wasn't as severe for HiFi-GAN, which is based on directly generating waveforms. 
Although BiVocoder is also based on DDSP, it overcomes the aforementioned issue. 
It is capable of extracting more compact long-frame-shift and low-dimensional features and faithfully reconstructing the waveform.

\begin{table}[t!]\Huge
	\centering
	\caption{MOS with 95\% confidence intervals and RTF of HiFi-GAN, Autovocoder and BiVocoder on the test set of the VCTK for TTS task. Here, ``$a\times$" represents $a$ times real time.}\label{tab2}{
		\adjustbox{width=0.47\textwidth}{
			\begin{tabular}{l c c c}
				\hline
				\hline
				&  {MOS$\uparrow$}& {RTF (GPU)$\downarrow$}& {RTF (CPU)$\downarrow$}\\
				\hline
								{Natural} &3.81$\pm$0.17&-&-\\
				\hline
				{HiFi-GAN} &3.78$\pm$0.20&0.00320 (313$\times$)&0.166 (6.02$\times$)\\
				
				{HiFi-GAN*} & 3.77$\pm$0.19&0.00315 (317$\times$)&0.162 (6.17$\times$) \\
				
				{Autovocoder} & 2.58$\pm$0.43&0.00169 (592$\times$)&0.00562 (178$\times$)\\
				
				{BiVocoder} & 3.77$\pm$0.19&0.00291 (344$\times$)&0.0368 (27.2$\times$)\\

				\hline
				\hline& 
	\end{tabular}}}
\end{table}

\vspace{-1mm}
\subsection{Evaluations on TTS Task}
\label{sec: TTS task}
\vspace{-1mm}

For the TTS task, STRAIGHT was excluded due to its poor performance in the analysis-synthesis experiments. 
Within the unidirectional neural vocoders, we only selected HiFi-GAN for TTS tasks because it exhibited better stability with different configurations of features as analyzed in Section \ref{sec: Analysis-synthesis task}. 
The DiffGAN-TTS\footnote{https://github.com/keonlee9420/DiffGAN-TTS.} was used as the acoustic model predicting mel spectrograms from text for HiFi-GAN, while for Autovocoder and BiVocoder, it predicted features extracted by themselves from the text. 
We also employed a speaker embedding model \cite{li2017deep} to assist the multi-speaker speech synthesis. 

The results of MOS subjective tests and RTF are shown in Table \ref{tab2}. 
It can be observed that using long-frame-shift and low-dimensional features in HiFi-GAN (i.e., HiFi-GAN*) achieved nearly identical MOS scores to the original HiFi-GAN. 
This further demonstrates the robustness of direct waveform prediction methods to feature configurations. 
Our proposed BiVocoder was comparable to HiFi-GAN and HiFi-GAN* in terms of synthesized speech quality, as indicated by the MOS results. 
However, it exhibited significantly higher waveform generation efficiency, particularly achieving around 4.5 times higher generation speed on CPU. 
This reflects the advantages of using DDSP-based methods, which are better suited for applications in resource-constrained scenarios, such as embedded devices. 
Unfortunately, despite both Autovocoder and BiVocoder belong to bidirectional neural network vocoders, Autovocoder with original feature configuration exhibited poor TTS performance, indicating that the features it extracted were difficult to predict. 
On the other hand, the features extracted by BiVocoder were acoustic-model-friendly and easier to capture.

\vspace{-1mm}
\subsection{Generalizability Validation}
\label{sec: Cross-dataests task}
\vspace{-1mm}

The bidirectional conventional vocoders (e.g., STRAIGHT) possess excellent generalizability, capable of feature extraction and waveform generation on any data without requiring additional data adaptation. 
To validate the generalizability of the comparative vocoders, we conducted analysis-synthesis experiments on the test set of the LJSpeech dataset. 
The trainable vocoders utilized a well-trained model on the VCTK dataset without further finetuning.
The experimental results are shown in the last column of Table \ref{tab1}. 
The BiVocoder still achieved the highest UTMOS score, confirming its strong generalizability for other data. 
Surprisingly, the Autovocoder achieved the sub-optimal results, indicating that the bidirectional neural vocoders had better generalizability compared to the unidirectional neural vocoders.

\vspace{-1mm}
\section{Conclusion}
\label{sec:conclu}
\vspace{-1mm}

In this paper, we have introduced a novel bidirectional neural vocoder called BiVocoder, which can not only extract long-frame-shift and low-dimensional features from waveforms but also reconstruct waveforms from these features. 
Experimental results demonstrated that for analysis-synthesis experiments, our proposed BiVocoder synthesized speech with higher quality compared to other vocoders and exhibited superior generalization across other datasets. 
TTS experiments demonstrated that the features extracted by BiVocoder were well-suited for prediction by acoustic models, achieving comparable results to baseline unidirectional neural vocoders, e.g., HiFi-GAN. 
Applying the BiVocoder to other speech generation tasks will be our future work.

\bibliographystyle{IEEEtran}
\bibliography{mybib}

\end{document}